\documentclass[%
 reprint,
 amsmath,amssymb,
 aps,
]{revtex4-2}

\usepackage{subfig}
\usepackage{graphicx}
\usepackage{dcolumn}
\usepackage{bm}

\begin{document}

\preprint{APS/123-QED}

\title{Towards long term sustainability of c-Si solar panels: the environmental benefits of glass sheet recovery}

\author{Marcos Paulo Belançon}
 \email{marcosbelancon@utfpr.edu.br}
\author{Marcelo Sandrini}%
\author{Francisnara Tonholi}
\affiliation{%
 Universidade Tecnológica Federal do Paraná (UTFPR), Câmpus Pato Branco, Brazil 
}%


\author{Leandro Silva Herculano}
\affiliation{
 Universidade Tecnológica Federal do Paraná (UTFPR), Câmpus Medianeira, Brazil
}%
\author{Gustavo Sanguino Dias}
\affiliation{Universidade Estadual de Maringá (UEM), Departamento de Física
}%


\date{\today}

\begin{abstract}
The cover glass in a silicon solar panel accounts for about 2/3 of the device's weight. Recycling these devices at their end-of-life is fundamental to reducing the industry's environmental impact. Here we investigate the recovery of these glass sheets by a heat-assisted mechanical process. A panel was delaminated, and we have utilized Fourier-transform infrared, Raman, and energy-dispersive spectroscopies to confirm the composition of the remaining components and identify aging signals. The results demonstrate that the panel's design was similar to most Silicon solar panels in the market, and we concluded that it would be feasible to recover the glass in most of these devices. Due to its chemical and mechanical strength, this glass would be ready to be reused without the need to melt it again, bringing substantial savings in its energy content and carbon emission related to its production. The glass sheet would be ready to be used as cover glass in another solar panel or architecture material. Our estimates showed that this could be a pathway to reducing the photovoltaic industry's carbon emissions by more than 2 million tonnes per year.
\end{abstract}

\maketitle


\section{\label{sec:level1}Introduction}

The development of crystalline silicon photovoltaic (c-Si) technology began in the 1950's \citep{Pearson1957}. However, it took decades until all the technology would be ready to be deployed and competitive in the world market. After the 2008 spike in fossil fuel prices, c-Si took the stage worldwide, promising to be a ``clean and affordable" alternative. In the 2010's, the photovoltaic (PV) installed capacity worldwide was multiplied by roughly 20, and industry projects it will reach the terawatt scale in the next few years \citep{Haegel2019}. The market share of c-Si is about 95 \% \citep{ITRPV2021}, and even though many other technologies are available \citep{Nayak2019}, none could rival c-Si production capacity. However, some scarce minerals, such as silver, may soon impose constraints \citep{LoPiano2019} on c-Si production growth \citep{Haegel2019}. Alternative materials and technologies are needed to reduce the raw material consumption of the solar electricity sector, such as efficient recycling  \citep{Feltrin2007, Tao2011, Graedel2015, Davidsson2017}.

On the other hand, enhancing the life-cycle of c-Si panels should reduce environmental impacts and carbon emissions, contributing further to our carbon budget and the related climate goals \citep{Bhandari2015, Pickard2017, Zhou2018, Peters2020}. The industry has pursued to reduce materials waste and replace expensive minerals with cheaper alternatives \citep{ITRPV2021}, which often contributes to reducing the environmental impacts. However, the need to upscale the PV production capacity keeps pressuring the demand for raw materials in the entire supply chain, offsetting industry efficiency gains.

The cover glass is the main component of c-Si solar panels by volume. At an average thickness of 3 mm \citep{ITRPV2021}, it accounts for about 7.5 kg/$m^2$, which demands massive industrial infrastructure to produce millions of glass sheets \citep{Burrows2015} per day to supply PV's industry. Additionally, bifacial c-Si panels \citep{Gu2020} are growing their market share worldwide, and analysts are expecting this emerging technology to dominate the market in the next decade~\citep{ITRPV2021}. This trend will boost the demand for flat glass production, an energy-intensive activity that emits significant amounts of carbon into the atmosphere. In Europe, for example, Schmitz et al. \citep{Schmitz2011a} have accounted for an average emission of 0.74 tons of CO$_2$ per ton of flat-glass produced. Such value can be significantly higher, depending on the heat sources used; however, as the glass composition is identical in any production line, one may estimate a baseline for the energy consumption and carbon emissions. The energy requirements to produce 1 kg of glass are between 2 to 3 kWh due to the high melting temperatures (1500-1600 $^o$C), while about 0.2 kg of carbon will be emitted as a result of heating the carbonates used as raw materials~\citep{Schmitz2011a}.

As the lifespan of most solar panels ranges between 20-30 years, the fast growth of this sector in the last decade means that in the 2030's, massive quantities of c-Si solar panels will reach their end-of-life (EOL). Researchers worldwide have been addressing this imminent challenge of recycling these devices \citep{Tao2015, Sica2018, Xu2018a, Deng2019, Mahmoudi2019a, Chowdhury2020}, which seems fundamental to recovering scarce minerals and enabling the multi-terawatt deployment of c-Si. At the same time, it could reduce energy demand, carbon emissions, and raw material extraction, contributing to the sustainability of this industry in the long term.

In this contribution, we investigate a route to drastically minimize the amount of waste from c-Si products through the recovery of glass sheets. Nevertheless, before discussing that, we present a brief review of today's management of c-Si waste.

\subsection{\label{sec:level2}The state of the art of PVs waste management}

From top to rear commercial c-Si panels are composed of glass, encapsulation material, c-Si cells (including interconnections), another sheet of encapsulation material, and a plastic back sheet~\citep{ITRPV2021} (typically made of Tedlar-Pet-Tedlar, or TPT). An aluminum frame is often the last part of finishing the panel, though some frameless panels are already on the market. The most common encapsulation material by far is ethylene-vinyl acetate (EVA), though the market share of alternatives such as polyolefins~\citep{Lopez-Escalante2016} is growing~\citep{ITRPV2021}.

It is pretty easy to remove the aluminum frame, and this component can be easily recycled. However, all the other parts are laminated together, resulting in a sealing fundamental to the panels to withstand the environmental conditions for several decades. Additionally, even though some high-value materials are present, such as silver or poly-crystalline Silicon, most of the panel's weight is plastics, aluminum, and glass, which make alone about 60-70 \% of the total weight~\citep{Xu2018}.

Researchers, governments, and industries are concerned about adequately managing this growing stream of EOL solar panels. The exact volume worldwide is not precisely known \citep{Mahmoudi2019a}, and Europe is the only continent with dedicated PVs recycling facilities in operation \citep{Heath2020}. Among the several routes available to process these products, some are good enough to allow the reuse of silicon solar cells; however, this method has been discouraged due to several drawbacks, such as unavoidable cracked wafers and the interest of industry in producing refurbished panels~\citep{Heath2020}.

Several delamination processes are available. The choice relies on several factors and may vary in some regions as the cost of inputs, such as energy, may be very different. In general, one may classify these methods as chemical, thermal or mechanical delamination~\citep{Deng2019}, while in some routes, these processes can be combined and applied at the same time~\citep{Chen2019, Lovato2021}. Every method has its positive and negative aspects, which may include high energy demand or the consumption and production of toxic chemicals~\citep{Chowdhury2020}. 

The glass sheet in c-Si PVs is often treated as a low value, recyclable, and less important part of the device. Indeed, if we compare the value of the minerals used in all parts of a PV panel, it is clear that metals and Silicon are by far the most expensive ones. However, the production of flat glass is an energy-intensive activity; by mass, the glass is a very significant part of a PV panel. Also, the glass sheet is a durable material that may endure much more than the PV's lifespan, and reusing it would be far more beneficial to the environment than recycling it.

In such a context, we have been searching for an environmentally friendly method that could enable the reuse of glass sheets. The technique should also be easily scalable to contribute to solving the challenging question of handling end-of-life c-Si panels.

\section{Materials and methods}

A small solar panel (Kyocera Solar, model KS20T) measuring 520x352 mm (0.18m$^2$), which was used for the last five years in an ilumination system in our university campus was selected to be mechanical delaminated. FTIR-ATR (Perkin Elmer Frontier), Raman (Bruker Senterra), and SEM/EDS (FEI Quanta-250/Oxford spectrometer model X-Act) were used to confirm the composition of the materials found in the panel. In order to delaminate the panel, we began by removing the aluminum frames, which was the easiest step. Next, we placed the panel in a stove preheated at 85 $^oC$ for half an hour to soften the EVA and reduce its adherence to the glass. At this point, we manually pulled the laminated layer containing the back sheet, Silicon, and the interconnections starting in the panel's border.

\section{Results}

\begin{figure}[h]
    \centering
    \subfloat[Degradation in the border of the panel]{\includegraphics[height=4.5cm]{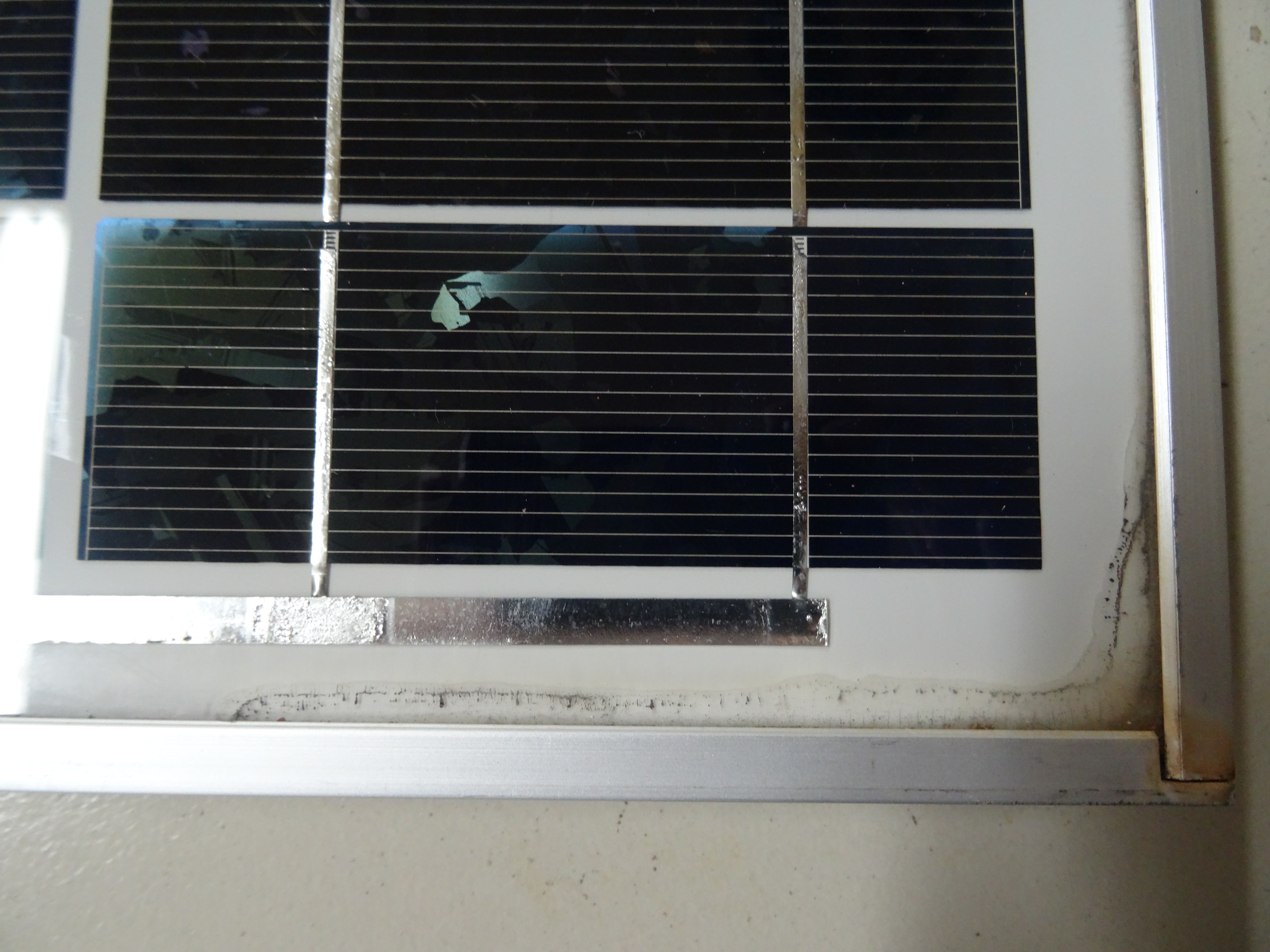}}
    \subfloat[Beginning of the mechanical delamination]{\includegraphics[height=4.5cm]{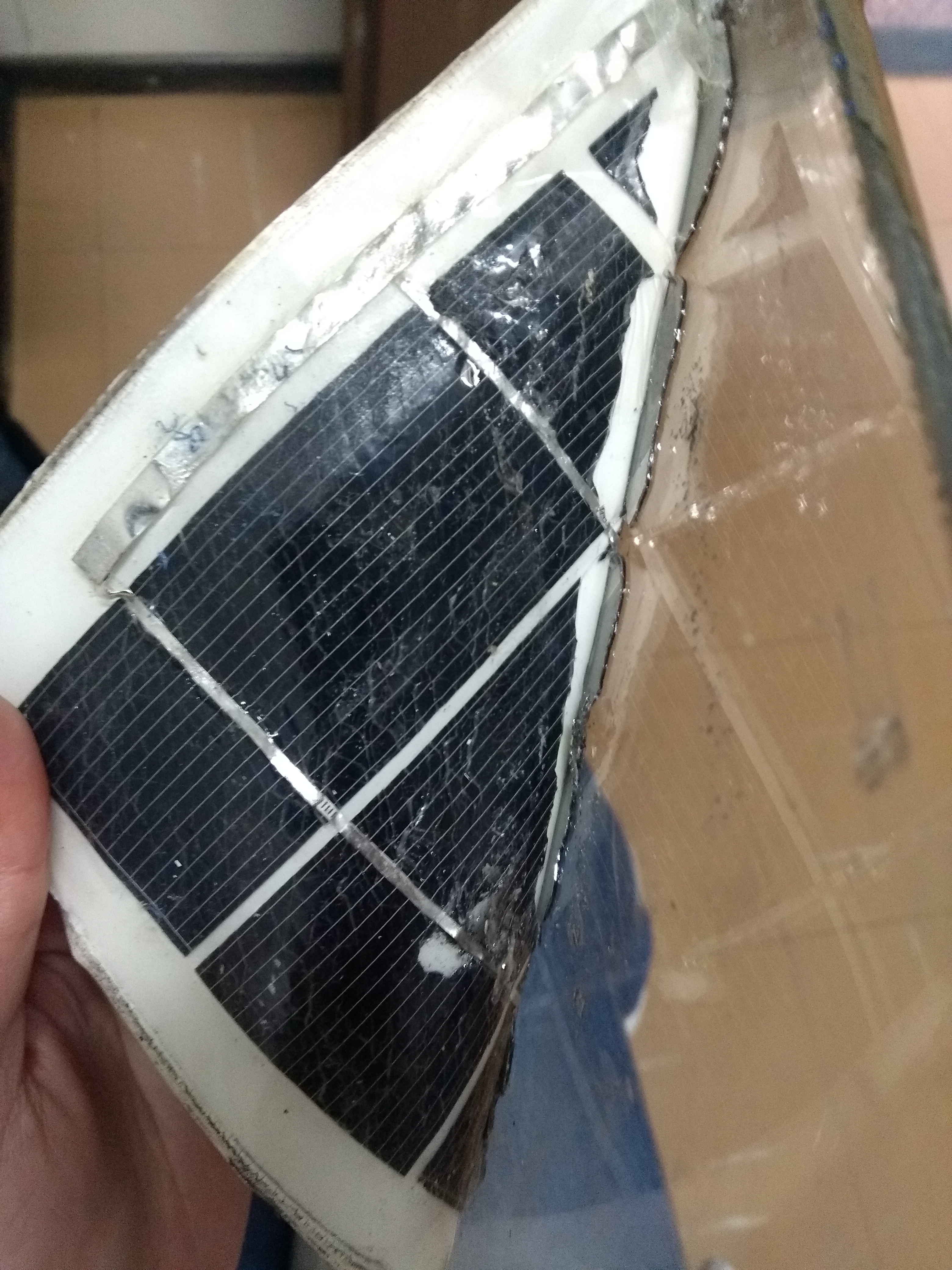}}
    \caption{Panel submitted to our delamination procedure}
    \label{fig:1}
\end{figure}

Figure \ref{fig:1} presents two photos showing regions of degradation (a) and the beginning of the delamination process (b).

\begin{figure}
    \centering
    \includegraphics[height=4cm]{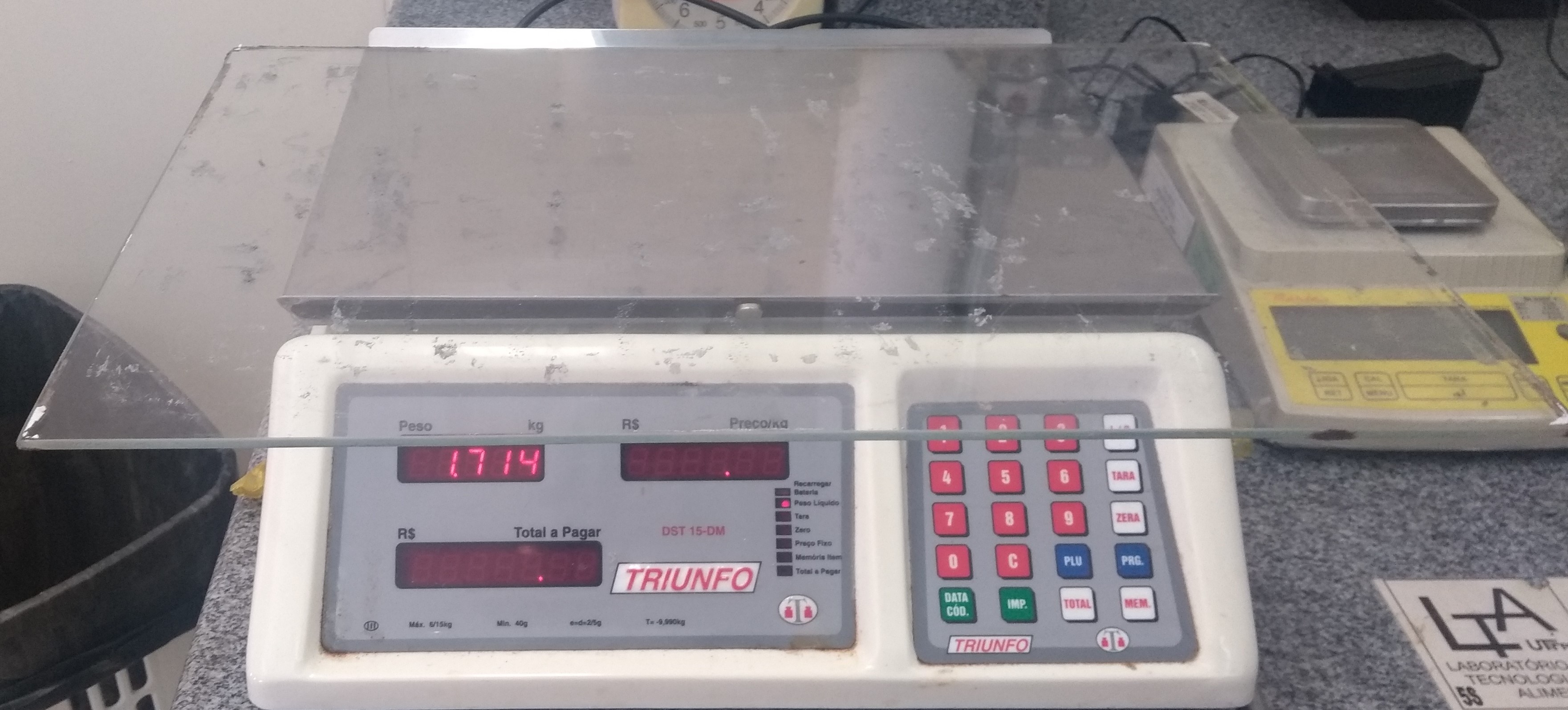}
    \caption{Recovered cover glass from the KS20T solar panel}
    \label{fig:2}
\end{figure}

 Thanks to the strength of the back sheet and the adhesion provided by the EVA lamination, we could separate the glass sheet. It still had some spots of EVA encapsulant at the end, which could include some points containing small quantities of Silicon. In figure \ref{fig:2}, one can see a photo of the recovered glass, and in table \ref{tab:1} the weight of the parts which were separated.

\begin{table}[h]
    \centering
    \begin{tabular}{llc}\hline
         Component & Weight (g)& Weight/Total (\%) \\ \hline\hline
         Glass&1714&71.5 \\
         Frame&372&15.5\\
         Others&308&12.8\\
         Total&2394&100\\\hline\hline
    \end{tabular}
    \caption{Weight of the solar panel used in this work.}
    \label{tab:1}
\end{table}

\begin{figure}[h]
    \centering
    \includegraphics[trim=100 0 0 0,width=10cm]{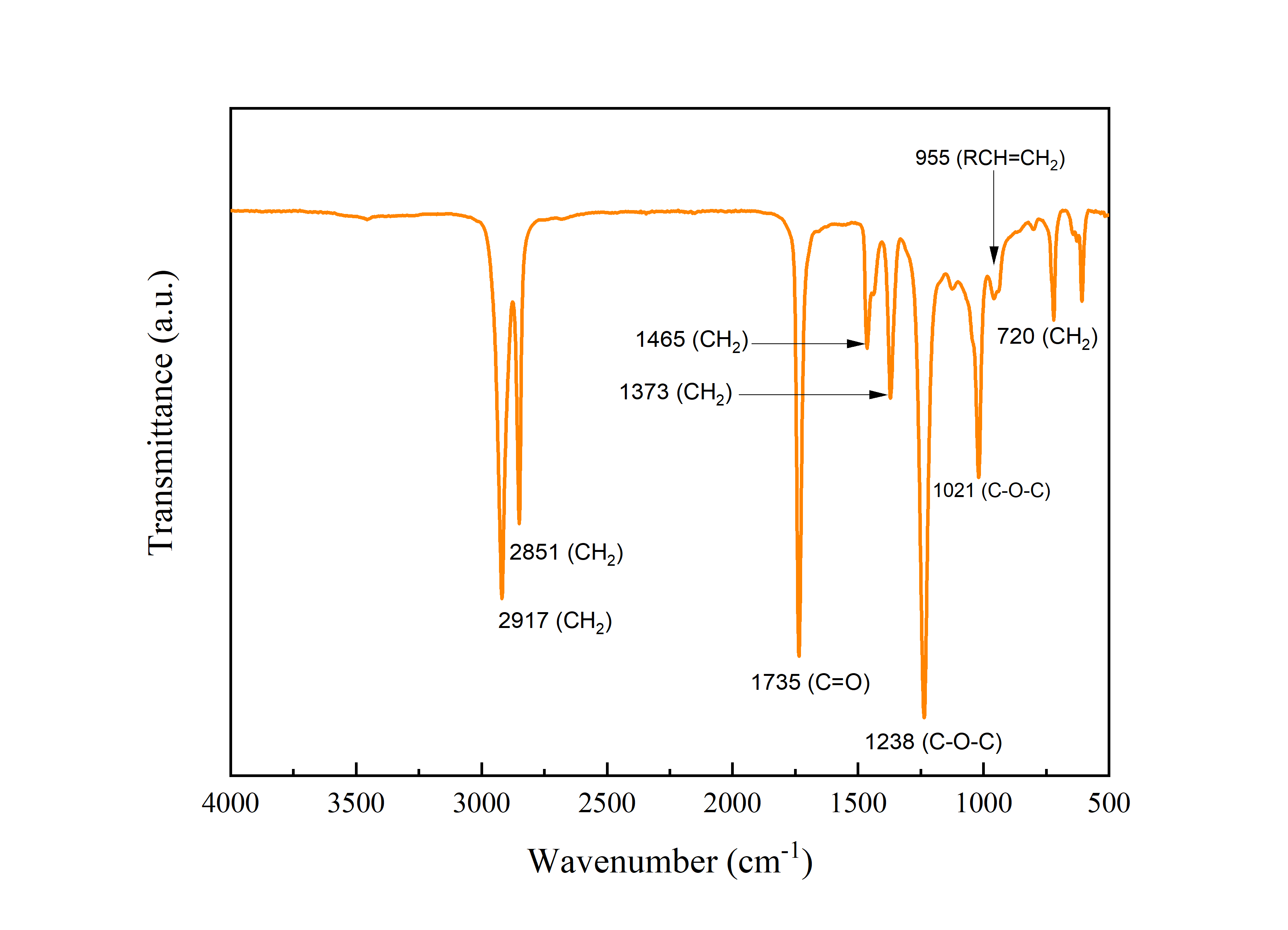}
    \caption{FTIR-ATR spectrum of the EVA from the KS20T panel, showing some characteristic lines~\citep{Marcilla2005,Jentsch2015}.}
    \label{fig:ftir}
\end{figure}

About 12-13 \% of the panel's original weight is ascribed as ``others" in table \ref{tab:1}. It consists of a sheet composed mainly of the back sheet and EVA, besides the metals and the Silicon, the most valuable parts. As one can see in figure \ref{fig:ftir}, FTIR-ATR measurements could identify several bands that are characteristic of the EVA encapsulant.

\begin{figure}[h]
    \centering
    \includegraphics[trim=1cm 0 0 0,width=10cm]{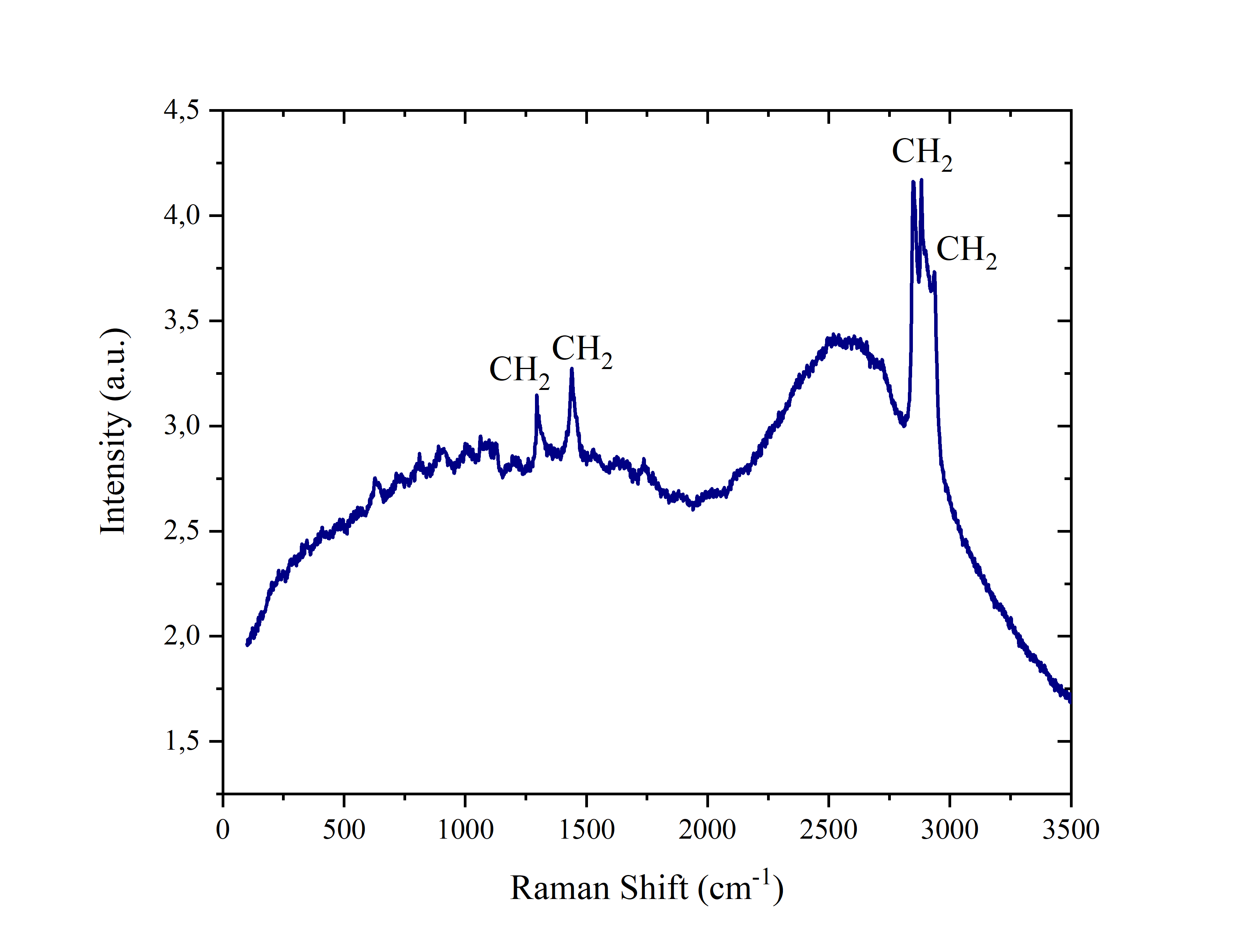}
    \caption{Raman spectrum of the EVA from the KS20T panel. Besides some characteristic lines~\citep{Peike2011a}, broad bands that we attributed to luminescence due to degradation can be observed~\citep{Schlothauer2012,Martinez2017}.}
    \label{fig:raman}
\end{figure}

Raman spectroscopy was performed under 532 nm pumping, and as one can see in figure \ref{fig:raman}, it also shows some characteristic lines from EVA. In addition, one can see an intense and broad emission that is attributed to the luminescence, as such result is similar to others found in the literature~\citep{Schlothauer2012, Martinez2017}. It is worth mentioning that several UV-blocking substances and other additives can be included in the EVA by the PV industry, and we do not know such details about the presence of this substances on the EVA in the KS20T panel. However, EVA luminescence has been used to quantify the degradation, as both functions of solar radiation dose~\citep{Jentsch2015} or penetration depth~\citep{Martinez2017}. As the KS20T panel had been exposed to the sun for several years, such luminescence under 532 nm pumping could be expected.

\begin{figure}[h]
    \centering
    \includegraphics[width=8cm]{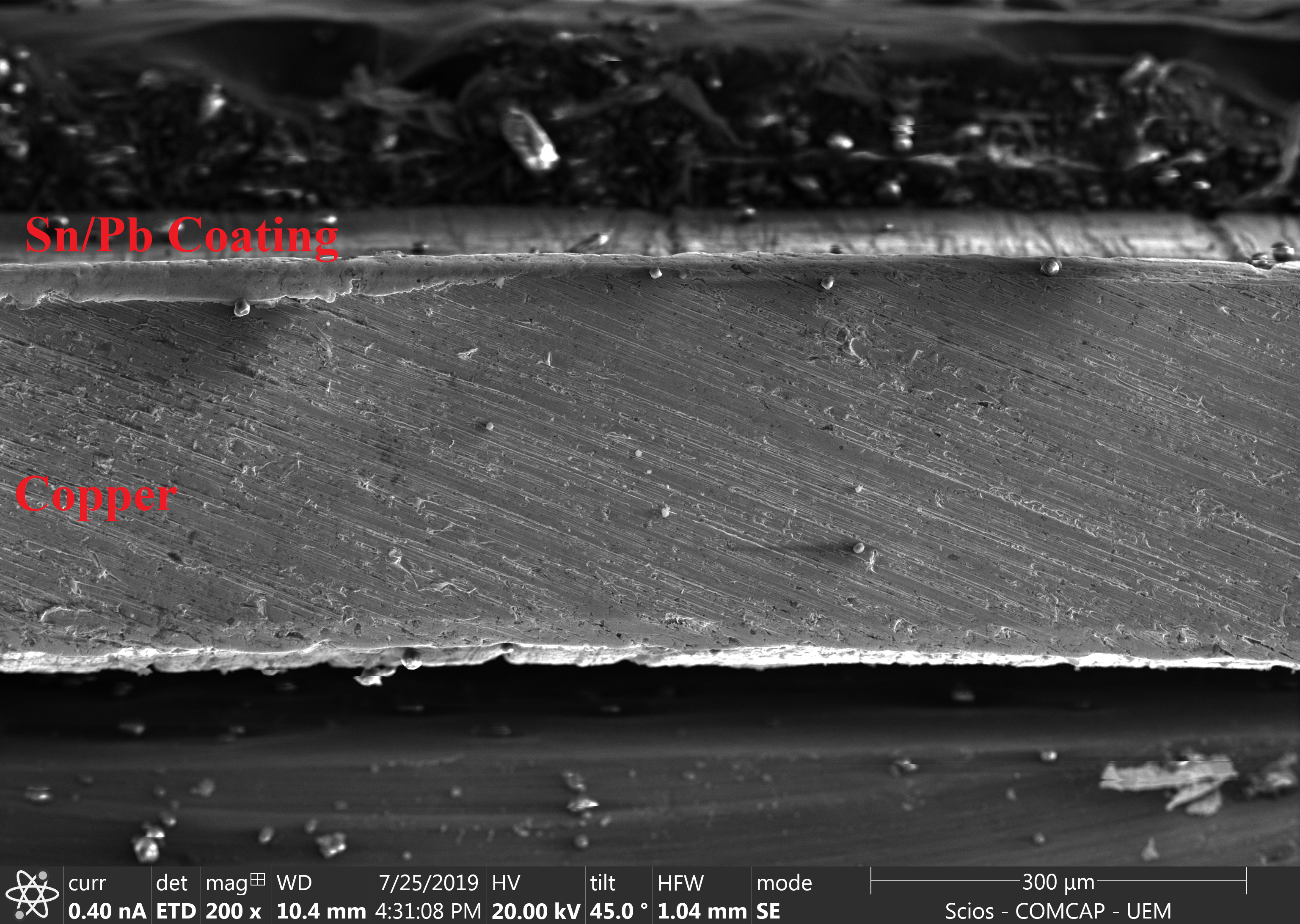}
    \includegraphics[width=8cm]{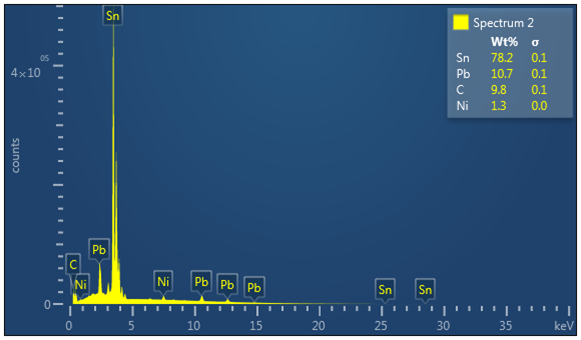}
    \caption{SEM image (up) and EDS spectrum (down) of an interconnection ribbon, demonstrating that we have the typical Sn/Pb coating.}
    \label{fig:eds}
\end{figure}

At this point, we had not investigated the EVA degradation enough, and we cannot say how it may have facilitated our manual delamination process of the KS20T panel. EDS measurements were performed in some pieces of the laminated sheet that remained, after the cover glass separation, to confirm the presence of several key elements. Besides Silicon, we could quickly identify Tin, Lead, Copper, and Silver. In figure \ref{fig:eds} we show an SEM image of a transversal section of an interconnection ribbon and the EDS spectrum of the coating observed.

From SEM images, we can see the thickness of the Sn/Pb coating is between 15 to 20 microns. Nevertheless, the industry expects that in the 2030's, most panels will use copper wires instead of ribbons~\citep{ITRPV2021}. Meanwhile, the reduction in volume/thickness of the materials in PVs indicates that the value of this remaining sheet may change and affect the choice of method to recover the metals in it.

\section{Discussion}

The manual removal of the cover glass was simple and worked as a proof of a concept. In theory, a machine could be developed to perform such a task, providing a pathway to process the massive amounts of PVs that will reach the EOL in the coming decades. Recovering Si cells for reuse is very delicate, and there is no demand for refurbished cells. However, it may be quite different in the case of the cover glass.

Glasses already have a high recycling rate in some countries~\citep{Jani2014}. However, even though a glass bottle has about the same composition as the cover glass in PVs, the latter has higher purity and transparency. In this way, it would be a loss to mix such high-quality material with colored glass bottles. Virtually all the solar panels in the market today have 60 or 72 Si cells, and their sizes remain between 1.8 m$^2$ and 2.2 m$^2$~\citep{ITRPV2021}. Soon, an industrial plant separating the cover glass from these panels would produce a continuous stream of high-quality flat glass, which could be reinserted in the PVs supply chain or, still, in another kind of application.

It is well-known that flat glass can be melted along with other raw materials and reduce consumption and emissions of glass production~\citep{Burrows2015}. Waste glass can also be used by cement and concrete industries~\citep{Jani2014}, and it can be recovered from windows and reused in buildings~\citep{Nubholz2019}. The possibility of having a stream of high-quality flat glass from EOL PVs, with sizes ranging around two m$^2$ (and 3 mm thickness), significantly enhances the glass value if compared to it in the form of ~15 kg of small pieces.

Buildings~\citep{Pariafsai2016} and architecture~\citep{Arbab2010} could widely explore such materials, adding value to the cover glass recovered from PVs and providing a pathway towards the development of a circular economy~\citep{Eberhardt2019, Nubholz2020}. To give some picture of the environmental benefits that it could bring to the PV industry, we made some estimates considering a standard solar module with 144 half-cell M6 (166x166mm$^2$), which has a size of about 2.2 m$^2$ and peak power of $\sim$450~W~\citep{ITRPV2021}. In table \ref{tab:2} we show the average weight and energy content of the cover glass in this kind of panel and the carbon emissions related to the production of this glass sheet.

\begin{table}[h]
    \centering
    \begin{tabular}{ccccc}\hline
         Panels&P(W$_p$)&Weight (kg)& E(kWh)& Emissions (t$_{CO_2}$) \\\hline\hline
         1&450&16.5&35.75&9.4$\cdot10^{-3}$\\
         244$\cdot10^6$&1.1$\cdot10^8$&4$\cdot10^9$&8.7$\cdot10^9$&2.3$\cdot10^6$\\\hline\hline
    \end{tabular}
    \caption{Average weight estimated from~\cite{ITRPV2021}, and energy content and emissions estimated by Schmitz et al~\citep{Schmitz2011a} for the European glass industry. The second row corresponds to about the total PVs produced in 2020.}
    \label{tab:2}
\end{table}

As one can see, there are several benefits to recovering the glass sheets. First, the energy content in all the cover glasses consumed by the PV industry in 2020 corresponds to about 8.7 TWh, equivalent to the annual electricity production of a 1 GW coal/nuclear power plant. Considering estimates for the European glass industry, where natural gas accounts for 80\% of the fuel consumed~\citep{Schmitz2011a}, the production of the glass covering the 110 GW$_p$ of PVs delivered last year~\citep{ITRPV2021} emitted more than 2 million tonnes of carbon into the atmosphere. Hu et al. ~\citep{Hu2018} have analyzed the emissions from container glass production in China, and their results indicated that such emissions could be ~50\% higher than in Europe due to the higher usage of Coal and fuel oil. As China is the biggest producer of solar panels today, the estimates made here may be considered conservative.

Among the alternatives, the approach we tested here has several advantages. For example, we have been investigating chemical delamination. We tested several solvents, such as Acetone, Acetic Acid Glacial, Hexane, Ethanol, Methyl isobutyl ketone, Isopropyl Alcohol, and Tetrahydrofuran. Only Tetrahydrofuran provided total delamination at room temperature and pressure, but it was necessary to smash the panel, destroying the glass entirely. Next, we cut the panel in small pieces (2x2cm) and immersed it in Tetrahydrofuran under stirring for a few hours. Such an approach is quite complex, and even though the chemical dissolution of EVA should increase significantly with temperature~\citep{Chen2019, Lovato2021}, its cost and environmental aspects may reduce its feasibility.

On the other hand, mechanical delamination often requires crushing, cutting, and smashing the panel~\citep{Tao2015, Guo2021}. This work has presented the alternative of recovering a c-Si cover glass unbroken by heating the panel to soften the encapsulant and mechanically separate the glass sheet. This approach enables several benefits due to the reuse of the glass and reduction in demand for new cover glass. As this material is responsible for about 60 to 70 \% of the panel´s weight, the method could severely reduce the amount of material submitted to other additional processes. The glass sheet recovery could enhance the environmental aspects of these other processes by reducing the demand for energy or chemicals if a thermal or chemical process is applied to recover the other components in the remaining sheet.

\section{Conclusion}
In summary, we could recover the cover glass of a 0.18 m$^2$ c-Si solar panel, and it seems feasible to develop some machine that could perform such mechanical delamination. FTIR and Raman measurements could detect some of the main characteristic lines of EVA. Raman under 532 nm pumping demonstrated some luminescence, interpreted as an aging signal. EDS confirmed the presence of several metals commonly found in c-Si panels. Our results indicate that this simple mechanical delamination could be scaled up to recover millions of cover glass sheets from EOL solar panels. Estimated energy conservation could range in 10 TWh due to the avoided demand for new cover glass production and, among other benefits, we estimate carbon savings equivalent to half the glass weight.

\begin{acknowledgments}
The authors wish to thank the Central de análises – UTFPR/PB and the COMCAP - UEM.
\end{acknowledgments}

\bibliography{library.bib}
\end{document}